\documentclass[english,aps,prl,superscriptaddress,floatfix,notitlepage,reprint,show pacs]{revtex4-2}
\usepackage[T1]{fontenc}
\usepackage[utf8]{inputenc}
\usepackage{physics}
\usepackage{natbib}
\usepackage{amsthm}
\usepackage{amssymb}
\usepackage{dsfont}
\usepackage{amsmath}
\usepackage{bm}
\makeatletter

\usepackage{braket}
\usepackage{txfonts} 
\usepackage{graphicx}

\usepackage[usenames,dvipsnames]{xcolor}
\usepackage[colorlinks=true,citecolor=Blue,linkcolor=RubineRed,urlcolor=Blue]{hyperref}
\usepackage{url}
\usepackage{tikz}
\usepackage{rotating} 
\date{\today}

\makeatother

\begin{document}
\title{ Different Phases in a Dissipative Rydberg Lattice : Roles of Occupancy and On-site Interaction }
\author{Suvechha Indu}
\address{Department of Physics, University of Calcutta, $92$ A. P. C. Road, Kolkata $700009$, India}
\email{suvechhaindu@gmail.com}
\author{Aniruddha Biswas}
\email{aniruddha.biswas@ltimindtree.com}
\address{Global Technology Office, LTIMindtree Limited, Kolkata, India}
\author{Raka Dasgupta}
\email{rdphy@caluniv.ac.in}
\address{Department of Physics, University of Calcutta, $92$ A. P. C. Road, Kolkata $700009$, India}
\begin{abstract}
We study a two-level dissipative non-equilibrium bosonic Rydberg system in an optical lattice, where multiple atoms can occupy a single site. The system is treated using two different approaches: solution of the master equation using a mean-field approximation, and direct numerical simulation of an equivalent quantum model. It is found that, depending on the on-site interaction strength, the system can either be uniform or have an antiferromagnet-like density-wave structure in terms of the Rydberg excitation distribution. Our mean-field treatment detects an interesting oscillatory phase as well, but the numerical simulation in 1D does not capture it. The origin of all these phases are investigated by studying the spatial correlations, and by calculating the fixed points of the dynamics. It is observed that an initial population difference across the sublattices helps to enhance the density-wave order. The scaling behavior of the system is also analyzed and a signature of weak universality is obtained. 

\end{abstract}


\maketitle
\section{Introduction}
Ultracold Rydberg atoms have attracted much attention in recent times because they serve as excellent platforms for quantum simulation \cite{morgado2021quantum, scholl2021quantum, kim2018detailed, surace2020lattice, weimer2021simulation, weimer2010rydberg, labuhn2016tunable, kim2024realization} and quantum information processing \cite{saffman2010quantum, lukin2001dipole, wu2021concise, cohen2021quantum, auger2017blueprint, ryabtsev2016spectroscopy}. Rydberg atoms, by virtue of having one electron excited to a very high principal quantum number, posses strong dipole moments. This dipole-dipole interaction among the atoms enforces a Rydberg blockade, where simultaneous excitations of nearby atoms are prohibited \cite{urban2009observation, gallagher2008dipole, gaetan2009observation, cote2006quantum}. The blockade mechanism can be harnessed to generate entangled states\cite{wilk2010entanglement, zhang2010deterministic, walker2008consequences, theis2016high, madjarov2020high, saffman2009efficient}. Rabi oscillations between the ground and excited states of Rydberg atoms, as well as the blockade, have been demonstrated experimentally \cite{isenhower2010demonstration, levine2018high, comparat2010dipole, gaetan2009observation, kubler2010coherent}.

The non-equilibrium properties of Rydberg systems are also being studied \cite{lee2011antiferromagnetic, hoening2014antiferromagnetic, carr2013preparation} in the presence of dissipation. This is very important because non-equilibrium features in many-body quantum systems often lead to interesting physical properties with no equilibrium counterparts \cite{carr2013preparation, lee2019coherent,polkovnikov2011colloquium, mondal2010non, gritsev2010scaling}, including newer aspects of criticality and universality. Now, in general, dissipation or noise in a Rydberg system destroys quantum coherence and is not favorable for quantum information protocols \cite{madjarov2020high}. However, it has been shown \cite{rao2013dark, chen2018accelerated, li2020periodically} that dissipation through spontaneous emission of atoms can actually facilitate the formation and stabilization of a two-particle entangled state \cite{carr2013preparation}. 

In \cite{lee2011antiferromagnetic}, an antiferromagnetic/density-wave order was predicted to emerge in a non-equilibrium Rydberg atom sequence, in terms of the Rydberg population distribution. Here, the atoms are fixed in space as in a lattice, and the lattice resembles a Rabi-coupled two-level system \cite{zeng1995nonclassical}. Activation of different phases can be done by varying either the Rabi frequency or the detuning parameter \cite{qian2012phase}. The antiferromagnetic phase here is essentially a consequence of the Rydberg blockade. 

In the present work, we consider an optical lattice where each site hosts multiple bosons, capable of being Rydberg excited. We study the nature of Rydberg excitations using mean-field theory, which is a good approximation for 2D or 3D systems. For a 1D system where mean-field theory fails, we use a full quantum simulation using a very small number of lattice sites. Thus, our approaches complement each other. We observe that there are certain important features common to the results obtained from mean-field and quantum simulation: the emergence of a density-wave order is the most crucial among them. 

Taking a mean-field assumption, we show that by tuning the on-site interaction, one can engineer two different phases : (i) a uniform or paramagnetic phase and (ii) a density-wave or antiferromagnetic phase in terms of the Rydberg excitation density. There is also an interesting subclass of the density-wave phase: an unstable oscillatory phase. The origins of these structures are probed by doing a fixed point analysis and also by studying the spatial correlations. Considering the on-site interaction to be the driving parameter, we construct the corresponding phase diagram. We study different aspects of this particular para-antiferro phase transition, including scaling forms near criticality. It is observed that this transition manifests weak universality with a linearly varying scaling exponent. We also show that if alternate sites of the lattice have an unequal number of atoms, then the Rydberg excitation distribution follows a density-wave pattern across a wider region in the parameter space. 

Using direct numerical simulation for a small 1D lattice, we find signatures of both the uniform and the density-wave phases. It is also observed that the spans of these phases depend on the dissipation coefficient. We compare these results with the mean-field solutions and obtain certain vital agreements. 

The paper is organized as follows. In Sec. II, we describe the model and derive the dynamical equations of motion using the mean-field approximation. In Sec. III A, the population dynamics is studied as a function of the on-site interaction; and in Sec. III B, different phases are characterized by defining a suitable order parameter. The spatial correlations of the system are probed using semi-classical Monte Carlo simulations in Sec. IV A, while the fixed points of the dynamics are investigated in Sec. IV B. In Sec. V, results from the direct numerical simulation are presented for a small lattice, and we compare them with the mean-field results obtained in Sec. III and Sec. IV. We study the critical behavior and scaling properties of the system in Sec. VI. The properties of the unequally populated sublattices are studied in Sec. VII. A brief summary is presented in Sec. VIII.

\section{Model Hamiltonian and Dynamical Equations}
Here, we investigate a two-level dissipative, non-equilibrium system with bosonic Rydberg atoms in an optical lattice. The system has multiple bosons occupying each potential minimum, and there are on-site interactions among them. So, in spirit, this is a Bose-Hubbard chain of Rydberg atoms in the zero-hopping limit.

Such a system can be created by trapping Rydberg atoms in an optical lattice. Rubidium atoms are excellent candidates for it because, on the one hand, there exist efficient techniques for trapping ultracold Rubidium in optical lattices \cite{low2007versatile, anderson2011trapping}, and, on the other hand,  Rubidium atoms can be excited to high principle quantum numbers and play the rolls of Rydberg atom \cite{thomas2018experimental,ramos2017measuring}. In an optical lattice, the low-hopping Mott insulator phase of a Bose-Hubbard model can be achieved experimentally \cite{greiner2002quantum, greiner2003ultracold} by increasing the depth of the optical lattice potential sufficiently ($20E_R$, where $E_R$ is the recoil energy). If it is increased up to $40 E_R$, that can be safely taken as the zero-hopping limit. 

In our model, if there are $N$ atoms in a site, they are collectively visualized as a ``superatom" as in \cite{petrosyan2013spatial}. Technically, each atom at the site $j$ can be in either of two states; one is the ground state $|g\rangle_j$, and the other is the Rydberg excited state $|e\rangle_j$.  There is spontaneous emission through which atoms from the Rydberg state can reach the ground state. Although any of the atoms in a site can in principle be Rydberg excited, a strong Rydberg blockade restricts multiple excitations from a single site, so it is either one excitation or none. If one atom from the $j$-th superatom reaches the Rydberg state, we consider that to be the excited state of the superatom $|E\rangle_j$. Similarly, when all constituent atoms are in the ground state $|g\rangle_j$, we call it the ground state of the $j$-th superatom $|G\rangle_j$. We now express the Hamiltonian in terms of the super-atomic states $|G\rangle_j$ and $|E\rangle_j$
\begin{equation}
    \begin{split}
        H  = & \sum_{j} \Big(-\Delta |E\rangle \langle E|_{j} + {{\Omega}\over{2}}(|E\rangle \langle G|_{j} + |G\rangle \langle E|_{j})\Big)\\ & + \sum_{j} {{U_{GG}}\over{2}}N_j(N_j-1) |G\rangle \langle G|_{j}\\&  + \sum_{j} {{U_{GG}}\over{2}}(N_j-1)(N_j-2) |E\rangle \langle E|_{j}\\ & + \sum_{j} U_{EG}(N_j-1) |E\rangle \langle E|_{j} \\& + V_0\sum_{<j k>} |E\rangle \langle E|_{j} \otimes |E\rangle \langle E|_{k}
    \end{split}
    \label{Hamiltonian}
\end{equation}

Here, $\Delta$ is the detuning between the laser frequency and the transition frequency among the atoms present at the same site, $\Omega$ is the Rabi coupling between the ground state and the excited state of the same site, and $V_0$ is the Rydberg interaction between the Rydberg excited atoms. In our work, $V_0$ is assumed to act among atoms that belong to nearest-neighbor sites only. Although the Rydberg interaction is long-range, the nearest-neighbor interaction is a valid approximation because the Rydberg interaction strength diminishes rapidly with an increasing separation. There are two on-site interaction parameters $U_{GG}$ and $U_{EG}$. $U_{GG}$ is acting among the ground state $|g\rangle_j$ bosons that occupy the same site $j$ and $U_{EG}$ is the interaction between the Rydberg excited boson and the ground state bosons, occupying the same site $j$. Since it is not possible to have two or more atoms in the excited state belonging to the same site, there is no such on-site term
for the excited state population. Thus, if the superatom is in the ground state, the on-site energy contribution is $(U_{GG}/2)N_j(N_j-1)$, $N_j$ being the net population of $j$-th site. If, on the other hand, the superatom is in the excited state, that means $(N-1)$ atoms are in the atomic ground state $|g\rangle_j$, and the on-site energy contribution is $(U_{GG}/2)(N_j-1)(N_j-2)$. In this case, the Rydberg excited atom interacts with the all $(N-1)$ ground state atoms with the energy $U_{EG}$ and hence the on-site interaction contribution here is $U_{EG}(N_j-1)$.

The time evolution of this system is described by a master equation, the usual framework for treating dissipative open quantum systems- \cite{overbeck2016time, weimer2021simulation,lee2019coherent, lee2011antiferromagnetic}.
\begin{eqnarray}
    \dot{\rho} = -i[H, \rho] + L[\rho]
\end{eqnarray}
where 
\begin{eqnarray}
    L[\rho] = \gamma \sum_j \Big(-{{1}\over{2}} \{|E \rangle \langle E|_j, \rho\} + |G\rangle \langle E|_j\rho|E\rangle \langle G|_j\Big) 
\end{eqnarray}

Here, $\gamma$ is the dissipation coefficient.

 We simplify the system using the mean-field approximation. This is done with the approximation: the Rydberg term $  |E\rangle \langle E|_j \otimes \sum_{k}|E\rangle \langle E|_k $ is approximated as $ |E\rangle \langle E|_j \otimes \sum_{k}\rho_{k,ee}$.
Here, $\rho$ is the density matrix of the system, with elements $\rho_{GG}, \rho_{GE}, \rho_{EG}$ and $\rho_{EE} $. We define $\omega_j = \rho_{j, EE} - \rho_{j, GG}$, and $q_j = \rho_{j, EG}$. Thus, $\omega_j$
essentially represents the state of a superatom at site $j$ (if it is in $|G\rangle$, $\omega_j=-1$, if it is in $|E\rangle$, $\omega_j=1)$. 

The time evolution equations now become: 

\begin{equation}
    \dot{\omega}_j = -2\Omega Im q_j - \gamma(\omega_j + 1)
\end{equation}

and
\begin{equation}
    \begin{split}
     \dot{q}_j = & i\Big[\Delta - {{V_0}\over{2}}\sum_{<j,k>}(\omega_k +1) + U (N_j - 1)\Big]q_j\\ & -{{\gamma}\over{2}}q_j + i{{\Omega}\over{2}}\omega_j
    \end{split}
    \label{u}
\end{equation}    

Here, $U=U_{GG}-U_{EG}$  is the relative on-site interaction parameter for our system. 
The term $(V_0/2)\sum_{<j,k>}(\omega_k +1)$ in the above equation can be written as $(zV_0/2)(\omega_{j+1} +1)$ where $z$ is the coordination number of the lattice, assuming isotropic interaction. 

It is observed that the system arranges itself in the form of a bipartite lattice, comprising sublattices 1 and 2. The system is analogous to a two-component spin system: a Rydberg excited superatom can be mapped to a down ($\downarrow$) spin, and a superatom in its ground state can be mapped to an up ($\uparrow$) spin.  If the value of $\omega_j$ is the same across both sublattices, the phase is termed as uniform, resembling a paramagnetic phase. If the average values of  $\omega_1$ and $\omega_2$ are different for the two sublattices, the phase is non-uniform, and resembles an antiferromagnetic order. 

Scaling all energies by $\gamma$, and time by $\gamma^{-1}$, and defining $V=zV_0/2$, we arrive at the following.

\begin{eqnarray}
\label{dynamics1}
   \dot{\omega}_1 &=& -2{\Omega} Im q_1 - \omega_1 - 1 \\ 
   \label{dynamics2}
   \dot{\omega}_2 &=& -2{\Omega} Im q_2 - \omega_2 - 1 \\ 
   \label{dynamics3}
   \dot{q}_1 &=& i[\Delta - V (\omega_2 +1) + U (N_1 - 1)]q_1 - {{q_1}\over{2}} + i{{\Omega}\over{2}}\omega_1 \\
   \label{dynamics4}
   \dot{q}_2 &=& i[\Delta - V (\omega_1 +1) + U (N_2 - 1)]q_2 - {{q_2}\over{2}} + i{{\Omega}\over{2}}\omega_2 
\end{eqnarray}

\section{Dynamics of the System}
\subsection{Uniform and Non-Uniform  Phases}
\begin{figure}
\begin{center}
\includegraphics[width=0.49\linewidth]{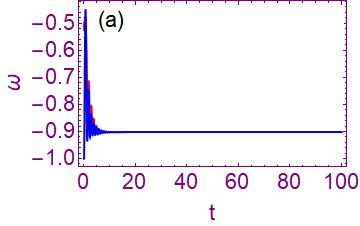}
\includegraphics[width=0.49\linewidth]{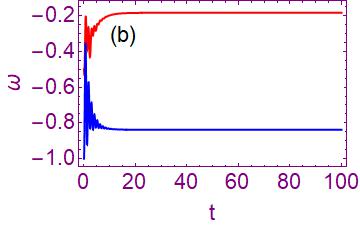}
\includegraphics[width=0.49\linewidth]{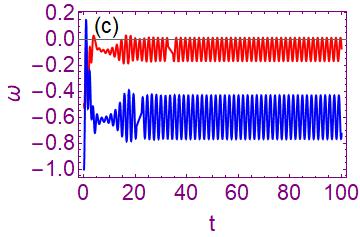}
\caption{Three phases, (a) Uniform Phase, $  \Delta = 1, \Omega = 3, V = 6, U = 2, N = 4$, (b) Antiferromagnetic Phase,$  \Delta = 1, \Omega = 3, V = 6, U = -0.3, N = 4$, (c) Oscillatory Phase, $  \Delta = 1, \Omega = 3, V = 6, U = 0.6, N = 4$ }
\label{threephases}
\end{center}
\end{figure}

We study the dynamics of the system by numerically solving Eq. \ref{dynamics1}-\ref{dynamics4}. For convenience, we take an equal number of atoms in both sublattices, i.e., $N_1 = N_2 = N$. It is seen that in the zero-hopping limit, the fluctuation in the atom distribution is zero, in that case the number of bosons occupied each site is equal to each other. So, an equal distribution of atoms in each site is achievable \cite{plimak2004occupation}. In Fig. \ref{threephases}, we present the $\omega$ vs. $t$  solutions using a chosen set of initial conditions ($q_1(0) = 0, q_2(0) = 0, \omega_1(0) = -0.5, \omega_2(0) = -1$), and for different parameter values. Here, the red curve (the upper line in Figs. \ref{threephases} (b) and (c)) represents sublattice 1, and the blue curve (the lower line in Figs. \ref{threephases} (b) and (c)) represents sublattice 2. We find two distinct phases in terms of the population distribution. In Fig. \ref{threephases} (a), the uniform phase is shown where both sublattices have equal and steady distributions, constant in time. Figs. \ref{threephases} (b) and (c) both represent non-uniform phases with density-wave structures \cite{yang2022density}. Here, (b) shows a steady population distribution pattern, and (c) shows a fluctuating population distribution. These are termed antiferromagnetic and oscillatory phases in \cite{lee2011antiferromagnetic}. A structure similar to the former is also termed antiferromagnetic in \cite{hoening2014antiferromagnetic, carr2013preparation}, while it has been called a density wave-ordered state in \cite{samajdar2021quantum, samajdar2020complex, yang2022density, lin2020quantum}. In this article, we use both the terms ``antiferromagnetic" and ``density-wave ordered" interchangeably to denote the phase in Fig. \ref{threephases}(b), though technically, it would be a perfect antiferromagnetic order in terms of the population distribution only if all the alternate sites are Rydberg excited. The dynamics in Fig. \ref{threephases}(c) is a subclass of the density-ordered phase where the distributions are not constant, but oscillate aperiodically.

\subsection{Order Parameter}
We construct an order parameter for this system to distinguish between the phases:
\begin{equation}
\theta = {{\lvert {\omega_1 - \omega_2} \rvert}\over{2}}   
\end{equation}

Thus, $\theta =0$ if both the sublattices have the same distribution of Rydberg atoms. If the system shows perfect antiferromagnetic order, i.e., alternate sites have Rydberg excitations, then $\theta =1$. For any configuration in-between, $\theta$ would have an intermediate value. 

\begin{figure}
\begin{center}
\includegraphics[width=0.49\linewidth]{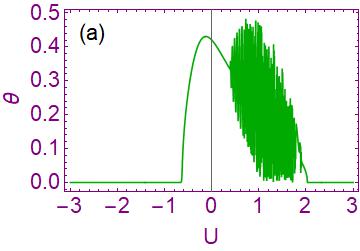}
\includegraphics[width=0.49\linewidth]{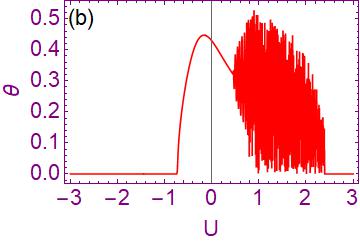}
\includegraphics[width=0.49\linewidth]{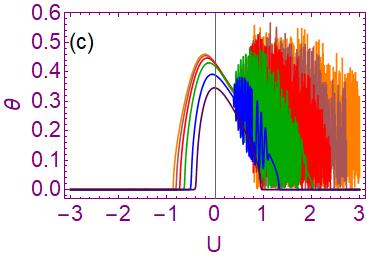}
\caption{(a) Order Parameter, $  \Delta = 1, \Omega = 3, N = 4, t = 100,  V = 8,$ (b) Order Parameter, $  \Delta = 1, \Omega = 3, N = 4, t = 100,  V = 10,$   (c) order parameter for various Rydberg interactions ($V = 5$ (purple), $6$(blue), $8$(dark green), $10$(red), $12$(dark pink), $14$(orange) and $ \Delta = 1, \Omega = 3, N = 4, t = 100$ )}
\label{op}
\end{center}
\end{figure}

In Fig. \ref{op} (a), we plot $\theta$ vs. $U$ for a particular time value $t$, keeping all other parameters fixed. We make certain observations: (i) there is a uniform phase at the extreme left; (ii) there is a non-uniform antiferromagnetic phase next; (iii) it is followed by another non-uniform but oscillatory phase; (iv) then again there is an antiferromagnetic phase; and (v) there is again a uniform phase at the right. 

However, in Fig. \ref{op} (b), with a higher Rydberg interaction, there is no antiferromagnetic phase on the right side, after the oscillatory phase.

In this plot \ref{op}(a), the zero values of $\theta$ represent the uniform phase, and  the non-zero $\theta$ region marks the non-uniform phase. In the non-uniform region, the single line $\theta$ part represents the constant antiferromagnetic phase, and the fluctuating $\theta$ part stands for the oscillatory phase.

We also plot the order parameter for various $V$ values in Fig. \ref{op}(c), and it is seen that as the Rydberg interaction increases, (i) the area of the non-uniform phases increases and (ii) $\theta_{m}$, i.e. the maximum value of $\theta$ that represents the highest possible amount of antiferromagnetic order, increases as well. 

\begin{figure}
\begin{center}
\includegraphics[width=0.49\linewidth]{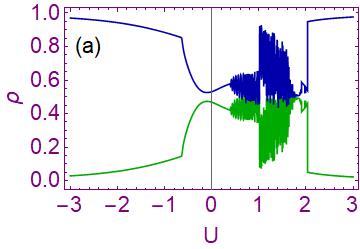}
\includegraphics[width=0.49\linewidth]{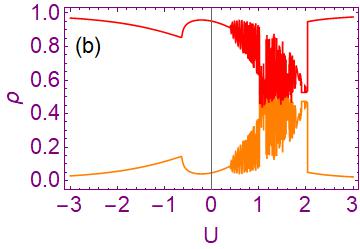}
\caption{The ground and the excited state populations for two sublattices, $  \Delta = 1, \Omega = 3, V = 8, N = 4, t = 100$, (a) blue and green colours represent the ground state and the excited state respectively of sublattice $1$ and (b) red and orange colours represent the ground state and the excited state respectively of sublattice $2$ }
\label{rho}
\end{center}
\end{figure}

The populations of the ground state ($\rho_{GG}$) and the excited state ($\rho_{EE}$) for the two sublattices under the mean-field approximation are shown in Fig.\ref{rho}. The excited state population is $\rho_{i,EE} = (1 + \omega_i)/2$ and the ground state population is $\rho_{i,GG} = (1 - \omega_i)/2$, here $i=1,2$ for two sublattices respectively.

It is clear from Figs. \ref{rho}(a)-(b) that, for large absolute values of the on-site interaction $U$, all atoms belong to the ground state. For small absolute values of $U$, some atoms go to the excited state, and the non-uniform phase emerges, as depicted in Fig. \ref{op}. 

\section{Origin of the phases}
In this section, we probe the origins of the non-uniform phases from two different perspectives: (i) by looking at the spatial correlation and (ii) by calculating the fixed points of the dynamics. 

\subsection{Studying Spatial Correlation}

 One way to characterize the phases in a more efficient way and also to investigate the origin of the three distinct types of particle dynamics, is to investigate the spatial correlations. However, since we are doing mean-field analysis, individual site-specific configurations are not known. We, therefore, do a semi-classical Monte Carlo simulation and generate a configuration of lattice sites that tallies with the mean-field results. 
For this, we first solve the dynamical equations (Eq. \ref{dynamics1} - Eq. \ref{dynamics4}) of the system to obtain the $\omega_1$ and $\omega_2$ values for each value of $U$, and hence  get the excited state population of the two sublattices. A random arrangement of lattice sites (i.e., whether the superatom is in $|G\rangle$ or $|E\rangle$) is generated that is consistent with the mean-field $\omega_1$ and $\omega_2$. In the language of spins, this would be equivalent to generating an Ising lattice with randomly allocated $\uparrow$ and $\downarrow$ spins with a pre-decided magnetization value.  

However, this array may contain two nearest-neighbor Rydberg excited states, which, in reality, are much less probable due to the Rydberg blockade. So, we define an additional transition probability and rearrange the sites slightly. To do this, first, we note that we can recast our Hamiltonian (Eq.\ref{Hamiltonian}) in a form without any explicit on-site interaction term. In this case, the on-site interaction $U$ is incorporated in a modified detuning term: 

\begin{equation}
    \begin{split}
        H_m  = & \sum_{j} \Big(-\Delta_m |E\rangle \langle E|_{j} + {{\Omega}\over{2}}(|E\rangle \langle G|_{j} + |G\rangle \langle E|_{j})\Big)\\ & + V_0\sum_{<j k>} |E\rangle \langle E|_{j} \otimes |E\rangle \langle E|_{k}
    \end{split}
    \label{Hamiltonianmodified}
\end{equation}

where, $\Delta_m = \Delta + U(N_j - 1)$. The Hamiltonians in Eq.  \ref{Hamiltonian} and Eq.  \ref{Hamiltonianmodified} both produce the same dynamics. 

The expression of the transition probability of a two-level Rydberg system with one atom per site is (\cite{petrosyan2013spatial}) ,
  \begin{equation*}
      P = {{\Omega^2}\over{(2\Omega^2 + {{\gamma^2}\over{4}} + (\Delta + V_0)^2)}}
  \end{equation*}

In our system, $\Delta$ would have to be replaced by $\Delta_m$, as there are multiple bosons per site. Considering this and rescaling the energy terms by $\gamma$ as before, we find the following. 

\begin{equation}
    P = {{\Omega^2}\over{(2\Omega^2 + {{1}\over{4}} + (\Delta_m + V)^2)}}
    \label{p}
\end{equation}

This is the probability that two consecutive sites would be Rydberg-excited. An increasing Rydberg interaction $V$ ensures a decreasing probability, as expected. 
So, the lattice configuration is marginally modified to ensure that the probability of having two nearby Rydberg excitations does not exceed $P$ in Eq.\ref{p}. 

Once the final configuration is obtained, we calculate the correlation $S_r$ between $j$-th and $(j+r)$-th sites, defined as:
 \begin{equation}
  S_r = \langle \omega_j \omega_{j+r} \rangle - \langle \omega_j\rangle \langle \omega_{j+r}\rangle   
 \end{equation}
 
\begin{figure}[th]
    \centering
\includegraphics[width=0.7\linewidth]{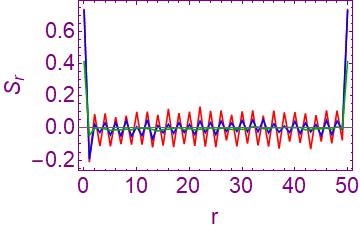}
    \caption{Correlations in the three phases, $\Delta = 1, \Omega = 3, V = 12, N = 4$ and $U = 0.3$ in the antiferromagnetic phase (red), $U = 0.5$ (blue) in the oscillatory phase and $U =  -1$ in the Uniform phase(green)}
    \label{cor1}
\end{figure}

\begin{figure}
\begin{center}
\includegraphics[width=0.49\linewidth]{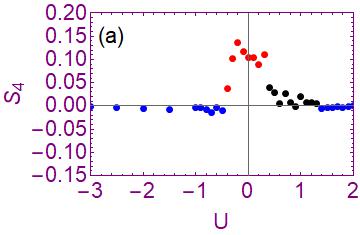}
\includegraphics[width=0.49\linewidth]{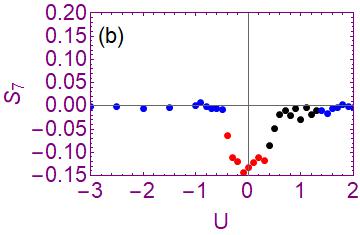}
\caption{The correlation with (a) even ($r=4$) site difference and (b) odd ($r=7$) site difference, here,  $  \Delta = 1, \Omega = 3, V = 12, N = 4$}
\label{cor2}
\end{center}
\end{figure}

  Because the value of $\omega_j$ depends on whether the $j$-th superatom is in $|G\rangle$ or $|E\rangle$, $S_r$ effectively measures how correlated the Rydberg excitations are. We plot $S_r$ as a function of $r$ for the three phases for a fixed $U$ in Fig. \ref{cor1}, and also for some fixed $r$ values as a function of the on-site interaction $U$ in Fig. \ref{cor2}. 

Fig. \ref{cor1} shows that the magnitude of the correlation is highest in the density-ordered phase (red curve). The zigzag pattern confirms that it is indeed an antiferromagnetic order. On the other hand, the correlation is lowest in the uniform phase (green) and is nearly zero. In the oscillatory phase (blue lines), the zigzag pattern is evident, too, so it is a subclass of the density-ordered phase. However, here, the correlation is far lower than the correlation in the steady antiferromagnetic phase due to fluctuations. Thus, the nature and magnitude of the correlation help to differentiate among the phases: the one with nearly zero correlation is the uniform phase; the one with strong antiferromagnetic correlation
is the density-wave phase; and the one with weak antiferromagnetic correlation is the oscillatory phase. This also explains the origin of the strange oscillatory phase: here, the excitations tend to occupy the sites alternatively, but the weak correlation forces the atoms to switch levels frequently. 

The simulation here is done with $50$ number of lattice sites in 1 dimension, averaging over $100$ configurations. We would like to point out here that in the mean-field equations (Eq. \ref{dynamics1} - \ref{dynamics4}) $V=zV_0/2$: thus a 2D system with Rydberg interaction $V_0$, and a 1D system with Rydberg interaction $2V_0$ yield the same phase plots, at the mean-field level. The mean-field results, of course, are reliable for higher-dimensional systems only, as the fluctuations from neighboring sites tend to cancel one another. We emphasize that by taking a 1D lattice for our semiclassical Monte Carlo, we do not intend to study an actual 1D system. Instead, a 2D system with Rydberg interaction $V_0$ is mapped to a 1D system with interaction $2V_0$ here because it keeps the dynamical equations unchanged. So, the nature of the correlations obtained here would actually be applicable for the 2D system and can be extended to 3D systems as well. 

We also plot $S_r$ with even site difference and odd site difference, respectively, with varying on-site interaction. For this, we choose two specific values of $r$: 4 and 7. 

In Figs. \ref{cor2}(a) and (b), $S_r$ vs. $U$ plots are presented for $r=4$ and $r=7$, respectively. The points belonging to antiferromagnetic phase, oscillatory phase, and uniform phase are denoted by red, black, and blue colors, respectively. It is evident that correlations in non-uniform phases (antiferromagnetic and oscillatory) are positive for even site difference (Fig.\ref{cor2}(a)) and negative for odd site difference (Fig. \ref{cor2}(b)).
The correlations in the uniform phase are close to zero, and there is no difference in having an odd site or even site difference in this particular range. We also observe that for any other even $r$ values, the plot resembles Fig.\ref{cor2}(a), and for any other odd $r$ values, the plot resembles Fig.\ref{cor2}(b). The correlation here does not diminish with increasing $r$, and no power-law/ exponential decay models of correlation would be applicable.

\subsection{Stability analysis of the fixed Points}

In the dynamical equations (Eq.\ref{dynamics1}-\ref{dynamics4}) $\omega_1$ and $\omega_2$ are real variables and $q_1, q_2$ can in principle be complex. Therefore, it essentially represents a set of 6 equations with $\Omega, \Delta, V, U$, $N_1, N_2$ as parameters. We calculate the fixed points of the system with the conditions $\dot{\omega}_1 = \dot{\omega}_2 = \dot{q}_1 = \dot{q}_2 = 0$. We obtain two classes of fixed points: a category with $\omega_1 = \omega_2$, and another category with $\omega_1 \neq \omega_2 $. There are three branches of the uniform fixed points from a cubic equation of $\omega$, and two branches of the non-uniform fixed points from a quadratic equation of $\omega$. 

We find that the uniform fixed points correspond to solutions of the following equation: 
\begin{equation}
    \begin{split}
         &V^2 \omega^3 - V(2 \Delta - 3V + 2 U (N - 1)) \omega^2 \\&  +  \Big({{\Omega^2}\over{2}} + {{1}\over{4}} + (\Delta - 3V) (\Delta - V) \\ & + U (N - 1) (2 \Delta -4V + U (N - 1))\Big)\omega \\ & + (\Delta - V)^2 + {{1}\over{4}} + U (N - 1) (2 \Delta -2V + U (N - 1)) =0      \end{split}
\end{equation}

The non-uniform fixed points are found to be the solutions of the following:

\begin{equation}
     \omega^2 - \omega^{\dagger} \omega + \Bar{\omega} =0
\end{equation}

Where,
\begin{equation}
    \begin{split}
       \omega^{\dagger} = & \Big({{1}\over{V + 4 \Delta^2 V + 2 \Omega^2 V + 4V U^2(N-1)^2 + 8\Delta VU(N-1)}}\Big) * \\& \Big(2\Delta + 8 \Delta^3 + 4\Delta \Omega^2 -2V \\& -  8V\Delta^2 -2V\Omega^2 + 8 U^3 (N-1)^3 \\& + U^2 (N-1)^2 (24\Delta - 8V) \\& + U(N-1)(24 \Delta^2 + 2 - 16 \Delta V + 4 \Omega^2)\Big)
    \end{split}
\end{equation}
and 
\begin{equation}
    \begin{split}
        \Bar{\omega} = & \Big({{1}\over{4V^2}} * ( 1 + 4 \Delta^2 \\& - 4V^2 + 2\Omega^2 + 4U^2(N-1)^2 \\& + 8 \Delta U (N-1))\Big) - \omega^\dagger
    \end{split}
\end{equation}

We vary the on-site interaction parameter $U$, and calculate the fixed points by solving the equations and finding the real roots. The corresponding stability properties are also investigated.

In the uniform phase on the left (Fig. \ref{op}(a)), there is only one stable branch of the uniform fixed points. In the antiferromagnetic region, the uniform fixed points become unstable, and there are two stable branches of the non-uniform fixed points. Interestingly, in the oscillatory region, no stable fixed points are found, and only unstable branches of the uniform and the non-uniform fixed points appear. That is why the Rydberg population is very fluctuating in this region. On the extreme right, the system is again in the uniform phase, where we have got two unstable branches of the non-uniform fixed points and three branches of the uniform fixed points, among which only one branch is stable and another two are unstable.

\begin{figure}
\begin{center}
\includegraphics[width=0.7\linewidth]{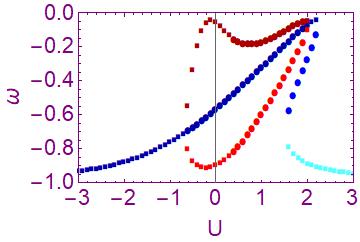}
\caption{Fixed Points for $\Delta = 1, \Omega = 3, V = 8, N = 4$. The stable uniform (non-uniform) fixed points are denoted by blue (red) squares and the unstable uniform (non-uniform) fixed points are denoted by blue (red) circles. }
\label{fp}
\end{center}
\end{figure}
In Fig. \ref{fp}, we have plotted the uniform and the non-uniform fixed points as a function of the on-site interaction $U$. The three branches of the uniform fixed points are denoted by three shades of blue and the two branches of the non-uniform fixed points are denoted by two shades of red. Here the stable fixed points are denoted by squares and the unstable fixed points are denoted by circles.

It is to be noted that, in absence of $\gamma$, no stable fixed point can be found. 

\section{Numerical simulation}

Mean-field calculations have proven to be very useful for a large number of quantum systems and quite dependable if the dimension is high and the system size is large. However, to bring out more intricate features of the dynamics, a full quantum solution is needed. If the system is 1D, mean-field theory fails miserably in general. These considerations prompt us to do a quantum simulation for a small lattice, and compare the qualitative results with the mean-field solutions.

Here, too, the modified Hamiltonian (Eq. \ref{Hamiltonianmodified}) is used, so everything is expressed in terms of $\Delta_m$, that contains $U$. The numerical simulation was conducted using Pulser \cite{silverio2022pulser}, an open-source Python-based package designed for Rydberg atom simulations by Pasqal. The simulation used a linear chain of eight sites, each representing a superatom, with the interatomic distance set equal to the Rydberg blockade radius of 9.0 $\mu$m. The key experimental parameters included a driving frequency $\Omega$ of 3 MHz and detunings $\Delta_m$ ranging from -5 to 5 MHz. The system was allowed to evolve for a total simulation time of 1000 ns under a depolarizing noise model, with noise strengths equivalent to decay rates (\( \gamma \)) of 0.25, 0.75, and 1.0 MHz. This setup provided an effective framework to study the influence of depolarizing noise on the Rydberg system's coherence and excitation dynamics. 

\begin{figure}
\begin{center}
\includegraphics[width=0.7\linewidth]{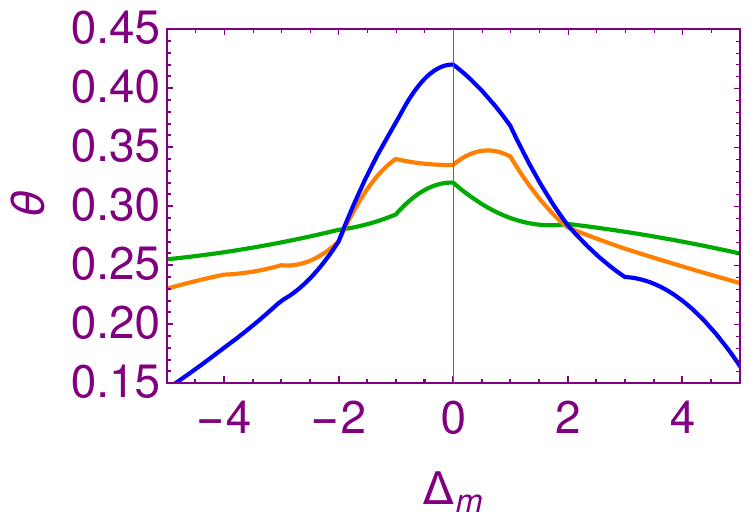}
\caption{$\theta$ vs $\Delta_m$ plot for $8$ sites with $V=16$, $\Omega=3$ and $\gamma=0.25$ (blue), $\gamma=0.75$ (orange), and $\gamma=1$ (green), as obtained from numerical simulation. }
\label{anitheta}
\end{center}
\end{figure}

\begin{figure}
\begin{center}
\includegraphics[width=0.7\linewidth]{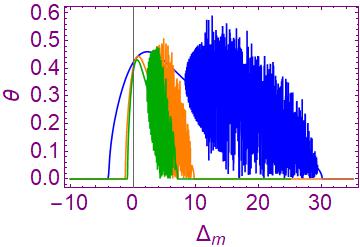}
\caption{ order parameter $\theta$ vs $\Delta_m$ plot, $\Omega = 3, V = 16 $, $\gamma = 0.25 $ (blue), $\gamma = 0.75 $ (orange), $\gamma = 1 $ (green), as obtained from Eq. \ref{dynamics1}-\ref{dynamics4}.}
\label{comparetheta}
\end{center}
\end{figure}

In Fig. \ref{anitheta}, the order parameter $\theta$ is plotted against the modified detuning $\Delta_m$ for a set of fixed $V$ and $\Omega$ values ($V = 16 $, $\Omega = 3$), and for different $\gamma$ values: 0.25, 0.75, and 1 (all in MHz, but we treat them as dimensionless parameters in subsequent parts because it is their ratios that matter). We calculate $\theta$ for a range of discrete $\Delta_m$ values, and fit it with interpolating polynomials. It is observed that $\theta_m$ shows a peak near $\Delta_m=0$ and falls with increasing $|\Delta_m|$. Clearly, the peaks signify antiferromagnetic order, and the tailing off signifies a gradual conversion to the uniform state. For the sake of comparison, we also keep a $\theta$ vs. $\Delta_m$ plot for the same set of parameters as obtained from the mean-field treatment of the dynamical equations (Fig. \ref{comparetheta}). We find that several key features are common in both phase plots: (i) there emerges a density-wave order centered near $\Delta_m=0$, (ii) a lower value of $\gamma$ leads to a wider range of this antiferromagnetic phase, and (iii) the highest value of $\theta_m$ is close to $0.4$ for this particular set of parameters. 
\begin{figure}[h]
\begin{center}
\includegraphics[width=0.7\linewidth]{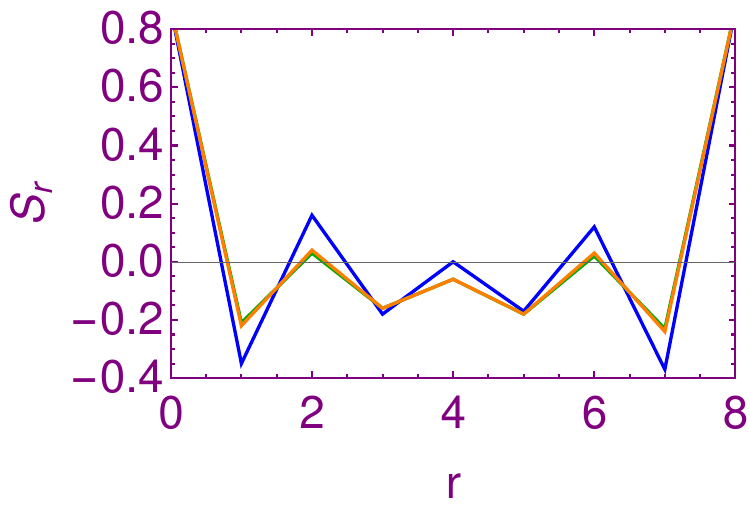}
\caption{$S_r$ vs $r$ plot for 8 sites, with $V=16$, $\Omega=3$ and $\Delta_m=0.04$, with $\gamma=0.25$(blue), $\gamma=0.75$ (orange), and $\gamma=1$ (green) as obtained from numerical simulations. }
\label{anicorsrvsr}
\end{center}
\end{figure}

\begin{figure}[h]
\begin{center}
\includegraphics[width=0.7\linewidth]{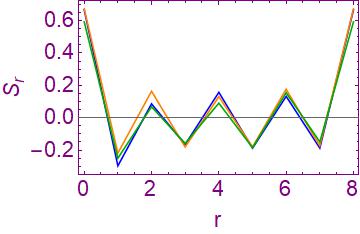}
\caption{ $S_r$ vs $r$ plot for $8$ sites, with $\Omega = 3, V = 16, \Delta = 0.04$, $\gamma = 0.25$ (blue), $\gamma = 0.75$ (orange), $\gamma = 1$ (green) obtained using Eq. \ref{dynamics1}-\ref{dynamics4}}
\label{comparecorr}
\end{center}
\end{figure}
There are certain aspects where the mean-field results deviate from the quantum simulation. For example, in the uniform phase, one achieves $\theta=0$ in mean-field, while in real simulations there would always be a non-zero $\theta$ value. Therefore, in contrast to the sharp antiferromagnetic transition observed in mean-field (which is applicable for higher-dimensional systems), the numerical simulation of a 1D system would show a smooth crossover from a state with a low antiferromagnetic order to a state with relatively higher antiferromagnetic order, as $U$ is varied. Also, the oscillatory region in the phase plot is not there in our numerical simulation: because one cannot capture its signature with such a small number of sites.  

The numerical simulation is also used to study spatial correlation of the system. We present $S_r$ vs. $r$ data for 8 atoms with periodic boundary conditions in  Fig. \ref{anicorsrvsr}, for $\Delta_m=0.04$ The zigzag pattern is a clear indication of the density-wave structure. The same is replicated using the dynamical equations in Fig.\ref{comparecorr}, and the qualitative nature matches fairly.  

\begin{figure}
\begin{center}
\includegraphics[width=0.49\linewidth]{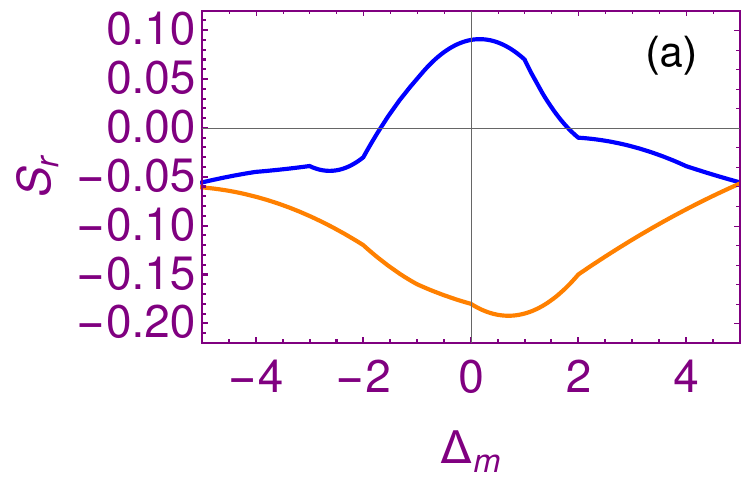}
\includegraphics[width=0.49\linewidth]{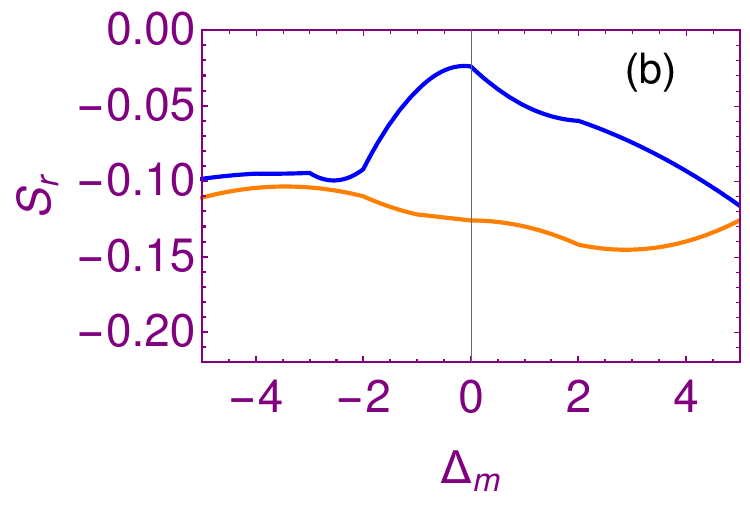}
\includegraphics[width=0.49\linewidth]{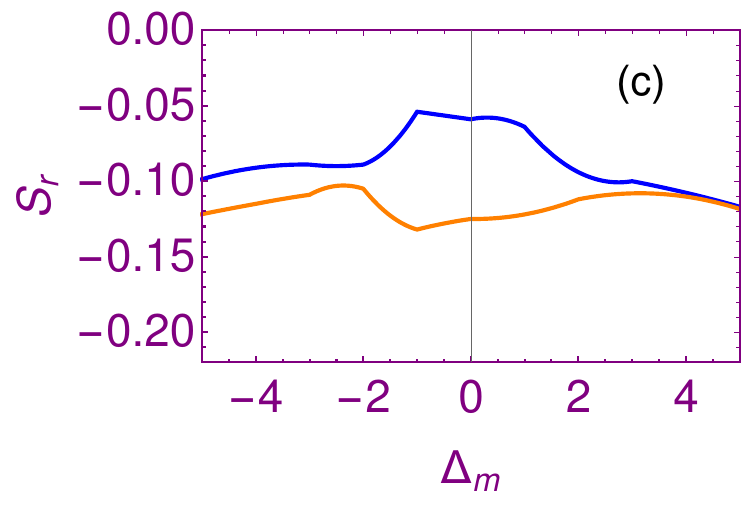}
\caption{ $S_r$ vs. $\Delta_m$ plots for $r=2$ (blue) and $r=3$ (orange). Here (a),(b), and (c) correspond to $\gamma$ values $0.25$, $0.75$, and $1$ respectively.}
\label{anicorsrdeltam}
\end{center}
\end{figure}

\begin{figure}
\begin{center}
\includegraphics[width=0.7\linewidth]{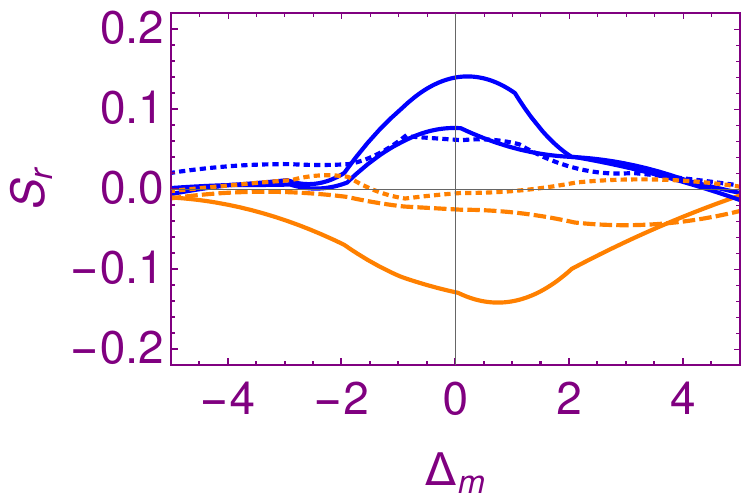}
\caption{ $S_r$ vs. $\Delta_m$ curves with baseline corrections. These are plotted for $r=2$(blue) and $r=3$(orange). The solid, dashed and dotted lines represent $\gamma=0.25$, $0.75$, and $1$ respectively. }
\label{anicorsrdeltamfull}
\end{center}
\end{figure}

\begin{figure}
\begin{center}
\includegraphics[width=0.49\linewidth]{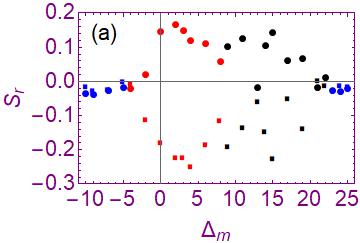}
\includegraphics[width=0.49\linewidth]{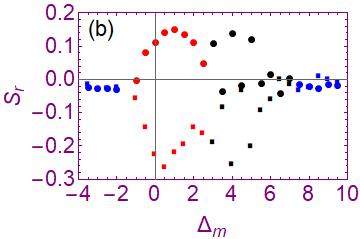}
\includegraphics[width=0.49\linewidth]{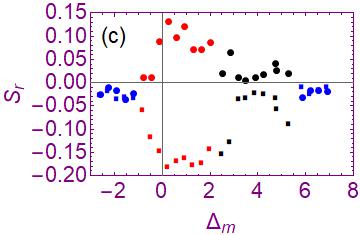}
\caption{ correlation $S_r$ vs $\Delta_m$ plot, $\Omega = 3, V = 16, $, (a) $\gamma = 0.25 $, (b) $\gamma = 0.75 $, (c) $\gamma = 1$. Even site difference ($r = 2$) is denoted by circles and odd site difference ($r = 3$) is denoted by squares. The antiferromagnetic phase, the oscillatory phase and the uniform phase denoted by red, black and blue colors respectively.}
\label{comparecor23}
\end{center}
\end{figure}

We also probe how $S_r$ depends on $\Delta_m$, so essentially on $U$. We plot $S_r$ against $\Delta_m$ for $r=2$ (blue) and $r=3$(orange) in Fig. \ref{anicorsrdeltam}. Here, Figs. \ref{anicorsrdeltam} (a), (b) and (c) correspond to $\gamma$ values $0.25$, $0.75$ and $1$, respectively. we find that the correlations for even $r$ and odd $r$ are nearly equal if $|\Delta_m|$ is large: hinting at a uniform state. For $\Delta_m=0$ and surrounding values, the correlation branches for odd-site and even-site differences maintain a gap: and it is thus an antiferromagnetic state. The gap is larger when $\gamma$ is small in magnitude. In an ideal case, the blue curve should always represent positive correlations, and the orange curve should always present negative values, similar to what one obtains from the mean-field results (Fig. \ref{comparecor23}). However, here we find that it is not always the case, and the correlation baseline (where the two branches meet) picks up non-zero values. This, probably, in an error induced by the small size of the lattice chosen due to computational constraints: the higher the value of $\gamma$, the larger the error is. We, therefore, do a baseline correction: shift the zeroes of the plots to the respective baselines, and replot (Fig. \ref{anicorsrdeltamfull}). Here the solid, dashed and dotted lines represent $\gamma=0.25$, $0.75$, and $1$ respectively. Blue curve represents $r=2$, and orange curve stands for $r=3$, as before. This shows that $S_r$ changes sign alternately, and the system certainly shows signatures of antiferromagnetism at low $|\Delta_m|$. 

\section{Critical behavior and scaling properties}

There are several works that address the question of critical phenomena in different variants of Rydberg systems. For example, if the complete long-range nature of Rydberg interaction is taken into account, it is observed to show critical exponents belonging to Ising universality class \cite{hoening2014antiferromagnetic, macedo2024fractonic}. For certain classes of Rydberg systems (such as with hard-core bosons \cite{ maceira2022conformal}, using the PXP model of Rydberg blockade \cite{menon2023emergent}, at multicritical points \cite{garcia2024numerical}), there are signatures of Ashkin-Teller criticality. In these works, either the detuning $\Delta$ or the Rabi frequency $\Omega$ or may be an external field (both the transverse and the longitudinal fields $g, h$, respectively \cite{wang2024quantum}) is the parameter that drives the phase transition. 

Our system is fundamentally different, as we only consider the nearest-neighbor Rydberg interaction (equivalent to saying that the blockade radius is equal to the lattice spacing in our model), and more importantly, it is the on-site interaction $U$ that serves as the driving parameter. We could not find traditional critical exponents here, but there are certain interesting scaling forms. 

\begin{figure}
\begin{center}
\includegraphics[width=0.49\linewidth]{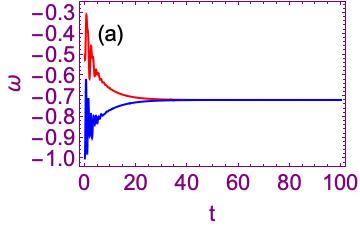}
\includegraphics[width=0.49\linewidth]{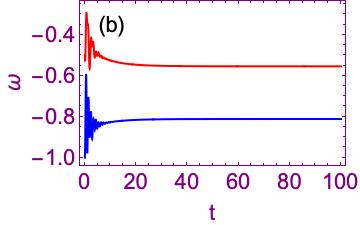}
\caption{Uniform phase to antiferromagnetic: continuous phase transition  $ \Delta = 1, \Omega = 3, V = 8, N = 4$.  (a) $U = -0.7 $ corresponds to the uniform phase just before the transition, (b) $U = -0.6$ corresponds to the antiferromagnetic phase, just after the transition.  }
\label{U1continuous}
\end{center}
\end{figure}

\begin{figure}
\begin{center}
\includegraphics[width=0.49\linewidth]{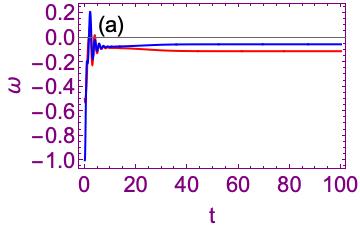}
\includegraphics[width=0.49\linewidth]{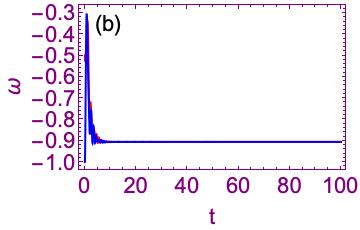}
\caption{antiferromagnetic phase to uniform:  discontinuous phase transition  $  \Delta = 1, \Omega = 3, V = 8, N = 4$. in (a) $U = 2 $ corresponds to the antiferromagnetic phase, just before the transition(b) $U = 2.1 $ corresponds to the uniform phase , just after the transition. }
\label{U2discontinuous}
\end{center}
\end{figure}

In our system, the uniform phase to the non-uniform phase transition (occurs on the left side of all the panels of Fig.\ref{op}) is a continuous transition, i.e., a second order phase transition. This is evident from Fig. \ref{U1continuous}, where $\omega_1$ and $\omega_2$, which overlap in the uniform phase (Fig. \ref{U1continuous}(a)), are gradually separating in the antiferromagnetic phase (Fig. \ref{U1continuous}(b)). On the other hand, the
non-uniform to uniform phase transition on the right side of Fig. \ref{op}(a) shows a sudden change of the order parameter, i.e., it is a discontinuous phase transition. This is evident from Fig.\ref{U2discontinuous}, where the values of $\omega_1$ and $\omega_2$ in the antiferromagnetic phase (Fig.\ref{U2discontinuous}(a)) do not merge continuously, to form the uniform phase (Fig. \ref{U2discontinuous}(b)). Therefore, the critical behavior is studied for the uniform to non-uniform transition at the left only.

\begin{figure}
\begin{center}
\includegraphics[width=0.49\linewidth]{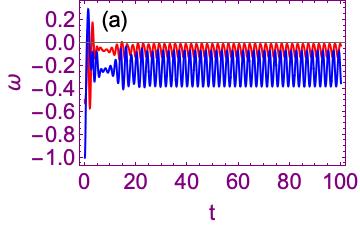}
\includegraphics[width=0.49\linewidth]{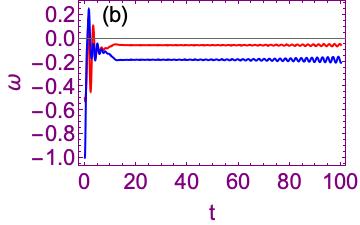}
\caption{Oscillatory phase to antiferromagnetic phase:  continuous phase transition  $  \Delta = 1, \Omega = 3, V = 8, N = 4$. in (a) $U = 1.8 $ corresponds to the oscillatory phase, just before the transition(b) $U = 1.9 $ corresponds to the antiferromagnetic phase , just after the transition. }
\label{oscafmcontinuous}
\end{center}
\end{figure}

As both the oscillatory phase and the constant antifferomagnetic phase are subclasses of the same non-uniform phase, they continuously evolve from one into another continuously. For example, the oscillatory to the antiferromagnetic crossover in Fig.\ref{op}(a) is a continuous one. This is reflected in Fig.\ref{oscafmcontinuous}, where $\omega_1$ and $\omega_2$, which are oscillating in the oscillatory phase (Fig \ref{oscafmcontinuous}(a)), gently go to the antiferromagnetic phase (Fig\ref{oscafmcontinuous}(b)), where 
the values of $\omega_1$ and $\omega_2$ in the antiferromagnetic phase (Fig. \ref{oscafmcontinuous}(b)) is close to
the mean values of the fluctuating $\omega_1$ and $\omega_2$ in the oscillatory phase (Fig.\ref{oscafmcontinuous}(a)) and where little oscillations still persists. 

In our system, as the Rydberg interaction increases, the antiferromagnetic phase on the right side of the phase diagram vanishes. It is evident from the right side of the Fig. \ref{op}(b), where the oscillatory phase directly transforms into the uniform phase via a discontinuous phase transition.


Let $( U_{m},\theta_{m})$ be the coordinate of the maximum antiferromagnetic order parameter and $U_{T}$ be the critical value of $U$, corresponding to the uniform to the antiferromagnetic phase transition at the left of the phase plots (Fig. \ref{op}). 
Our system shows scaling forms near both the critical point $U_{T}$ and the maximum antiferromagnetic point $U_{m}$.

To explain the scaling laws, let us define the parameters $\widetilde{\theta} = (\theta / \theta_m)$, $u_1 = |(U - U_{T})/ U_{T}| $ and $u_2 = |(U - U_T)/(U_m - U_T)|$.  

\subsection{Near critical point $U_{T}$}
Here, we study the behavior of our system near the critical point $U_{T}$, where, from the uniform to the antiferromagnetic phase transition occurs (Fig.\ref{op}).
It is empirically found that there is a scaling relation between $u_1$ and $\widetilde{\theta}$.

\begin{equation}
    \widetilde{\theta}^{1/\beta} + \Big({{u_1 - u_{1m}}\over{u_{1m}}}\Big)^2 = 1
    \label{eqbeta}
\end{equation}
Where, $u_{1m} = |(U_{m} - U_{T})/ U_{T}| $.

\begin{figure}
\begin{center}
\includegraphics[width=0.7\linewidth]{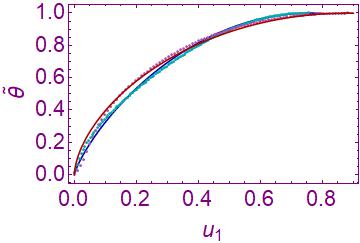}
\caption{ $\widetilde{\theta}$ vs. $u_1$ plot, and the curve obeying the relation \ref{eqbeta} for, $  \Delta = 1, \Omega = 3, N = 4, V = 6 $(red), $ = 14 $(blue)}
\label{twofit}
\end{center}
\end{figure}

\begin{figure}
\begin{center}
\includegraphics[width=0.7\linewidth]{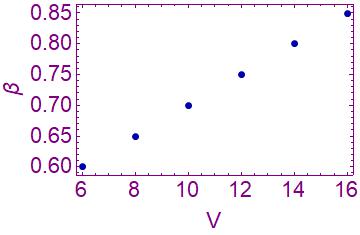}
\caption{The exponent $\beta$ vs. $V$ plot , $  \Delta = 1, \Omega = 3, N = 4$}
\label{coefeqn}
\end{center}
\end{figure}

This scaling form holds throughout the entire range of $u_1$ considered, except at a few points in the very close vicinity of $(u_1, \widetilde{\theta}) = (0, 0)$ (Fig. \ref{twofit}, where the red-colored curve corresponds to $ V = 6$ and the blue curve corresponds to $V = 14$, keeping other parameters fixed at $ \Delta = 1, \Omega = 3, N = 4$). This departure is more prominent for lower values of $V$, and is negligible for stronger Rydberg interactions. Here, the coefficient $\beta$ is not a constant, but a function of the Rydberg interaction strength $V$. We plot $\beta$ vs. $V$ for various Rydberg interactions (Fig.\ref{coefeqn}).

Here, it is seen that (Fig. \ref{coefeqn}) $\beta$ increases linearly as $V$ increases. It is found that $ \beta = 0.025 V+ 0.45$

For small $u_1$, when the system is near the critical point, i.e. near $U_{T}$, neglecting the $u_1^2$ term in the Eq. \ref{eqbeta}, we find that $\widetilde{\theta}$ varies as $u_1^{\beta}$ and we can write $\widetilde{\theta} = c_1 u_1^{\beta}$. Here, $c_1 = (2/u_{1m})^{\beta}$ is a constant, but its value depends on the Rydberg interaction strength $V$. The coefficient $\beta$ here, in a sense, plays the role of a traditional critical exponent. However, as Eq. \ref{eqbeta} is not very accurate in a small segment near $u_1=0$, and this notation of the critical exponent holds when $u_1$ is small, but not exactly near $u_1=0$.  

\begin{figure}
\begin{center}
\includegraphics[width=0.7\linewidth]{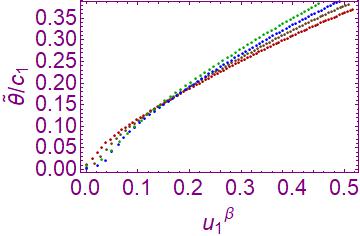}
\caption{The exponent $\widetilde{\theta}/c_1$ vs. $u_1^{\beta}$ plot , $  \Delta = 1, \Omega = 3, N = 4, V = 10$ (green), $V = 12$ (blue), $V = 14 $(brown), $V = 16$ (maroon)}
\label{UT}
\end{center}
\end{figure}

Again, $\widetilde{\theta}/{c_1}  = u_1^{\beta}$, so the curve of $\widetilde{\theta}/c_1$ vs. $u_1^{\beta}$ is a straight line that crosses the origin. We plot these curves for $V = 10 $ to $V = 16$ (Fig.\ref{UT}).

It is seen in the Fig. \ref{UT}, that, near $U_T$, i.e. for small $u_1$ or $u_1^{\beta}$, the curves overlap with each other and have the shape of a zero crossing straight line. However, very close to the origin (which corresponds to $u_1^{\beta} < 0.1, u_1 < 0.04$), they deviate nominally, as in Eq. \ref{eqbeta} is less accurate in that small window. So, if the limit $u_1\rightarrow 0$ is discounted, then $\beta$ serves as an exponent in the low $u_1$ regime. 

\subsection{Near maximum antiferromagnetic point $(U_{m})$}
Here, we study the scaling behavior near maximum antiferromagnetic point.
We find that there is a scaling relation that describes how the $\widetilde\theta$ vs. $u_2$ curve in the antiferromagnetic region behaves:
\begin{equation}
     \widetilde{\theta}=(1-u_2^2)^n
    \label{scaling}
\end{equation}

\begin{figure}
\begin{center}
\includegraphics[width=0.7\linewidth]{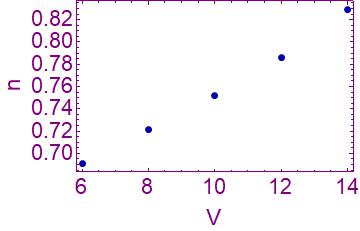}
\caption{The exponent $n$ vs. $V$ plot , $  \Delta = 1, \Omega = 3, N = 4$}
\label{exponent}
\end{center}
\end{figure}

Here, $n$ is not a constant, but varies linearly(Fig. \ref{exponent}) with the Rydberg interaction strength. 

We extract the $n$ values for different $V$ strengths, from the curves shown in Fig.\ref{op}(c). It is found that $ n =0.017 V +0.585 $ for our set of $\Delta$ and $\Omega$ (Fig. \ref{exponent}). 
When the order parameter is near $U_{m}$, i.e. for small $u_2 $ one can approximate Eq. \ref{scaling}, neglecting the higher order terms of $u_2$ as $\widetilde{\theta}\approx 1-n u_2^2$. This can be also be expressed as 
 $f1(\theta)\approx  {(1 - \widetilde{\theta})}/({0.017 V +0.585})  = u_2^2$

\begin{figure}
\begin{center}
\includegraphics[width=0.7\linewidth]{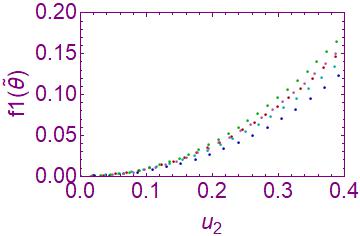}
\caption{Scaling form of $f1(\theta)$ vs. $\widetilde{U}$ with the parameters $\Delta = 1, \Omega = 3, N = 4$, $V$ = 6 (dark blue), 8 (cyan), 10 (red), 12 (purple), 14 (green).}
\label{um}
\end{center}
\end{figure}

This is shown in Fig. \ref{um}, where $f1(\theta)$ vs. $u_2$ plots are presented for fixed $\Delta$ and $\Omega$; and a range of $V$ values. It is observed that the points collapse on a single parabola near $u_2=0$, i.e., near $U=U_m$. The curves start to deviate as one moves farther away from $u_2=0$.

Neither $\beta$ nor $n$, are critical exponents in the conventional sense. The power-law scaling form involving $n$ is applicable close to the maximum antiferromagnetic point $U_m$, and not the critical point. In contrast, a scaling form involving $\beta$ is valid near the critical point $U_T$, but does not address the exact limit $U \rightarrow U_T$. Thus, none of them qualifies as traditional critical exponents. However, both $n$ and $\beta$ help to describe the system in a scale-free form, in two opposing limits, and play the roles of effective scaling exponents. As both $\beta$ and $n$ vary linearly with the Rydberg interaction strength $V$, the system might belong to a weak universality class, as reported in other statistical systems \cite{khan2017continuously, sau2023weak,furlan2019weak}. It might also be a signature of hidden superuniversality as in \cite{mukherjee2023hidden}.

\section{Unequally populated sublattices}
So far, we have only discussed sublattices populated equally, where odd-numbered and even-numbered sites contain the same number of atoms. 
Therefore, the superatoms on each of these sites are identical : they comprise an equal number of atoms. There is another possibility through selective loading: a population difference can be incorporated in the two sublattices, so that the superatoms in the two sublattices are different in terms of number of constituent atoms. This selective loading can be achieved by generating a long-period optical lattice and a short-period optical lattice in two different planes and applying the long- and the short-period lattice sequentially
\cite{aidelsburger2013realization, aidelsburger2011experimental, peil2003patterned, schreiber2015observation}. To study this system, we choose $N_1 \neq N_2$ in Eqs. \ref{dynamics3}, \ref{dynamics4}  and construct the $\theta$ vs. $U$ phase plot. We find that instead of the sharp uniform to non-uniform transition  in the $N_1 = N_2$ case, here the $\theta$ vs. $U$ profile changes slowly and resembles a bell-shaped curve.

\begin{figure}
\begin{center}
\includegraphics[width=0.49\linewidth]{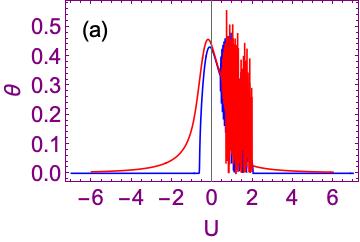}
\includegraphics[width=0.49\linewidth]{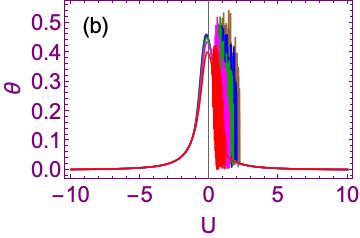}
\caption{The order parameter $\theta$ vs. on-site interaction $U$ plot for (a)$  \Delta = 1, \Omega = 3, V = 8,  N_1 = 4, N_2=6 $ (red), and $N_1 =  N_2 = 4$(blue) (b) $  \Delta = 1, \Omega = 3, N_1 = 4, N_2=6 $, V = 6 (red), V = 8 (magenta), V = 10 (green), V = 12 (blue), V = 14 (brown)}
\label{Inequal}
\end{center}
\end{figure}

Here in Fig. \ref{Inequal}(a), the red curve represents the $\theta$ vs. $U$ plot for unequally populated  sublattices. The antiferromagnetic region here is larger than its $N_1=N_2$ counterpart, represented by the blue curve. Even a small difference in the population of the two sublattices leads to an increase in the width of the non-uniform phase along the $U$-axis. However, if that imbalance increases further, the width remains unchanged, and the peak value of $\theta$  increases marginally. 
We also plot the phase diagrams for unequally populated sublattices with different Rydberg interaction strengths, and it is seen that (Fig. \ref{Inequal}(b)) the critical point $U_{T}$ for each curve, where from the uniform to the antiferromagnetic phase transition occurs, has the same value. 


Here, in the phase diagram of the unequally populated sublattices, $\widetilde{\theta}$ exponentially rises from the uniform phase to the antiferromagnetic phase (Fig. \ref{Inequal}(b)) as $U$ is varied.
This exponential increase can be expressed in terms of $\widetilde{\theta}$ and $u_1$ as
\begin{equation}
    \widetilde{\theta}  =e^{\kappa(u_1 - u_{1m})}
    \label{kappa}
\end{equation}

\begin{figure}
\begin{center}
\includegraphics[width=0.7\linewidth]{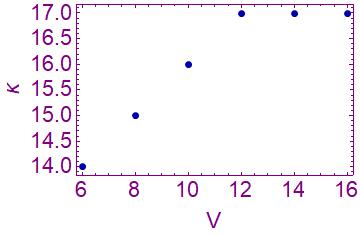}
\caption{The exponent $\kappa$ vs. $V$ plot , $  \Delta = 1, \Omega = 3, N_1 = 4, N_2 = 6$}
\label{coefnoteqn}
\end{center}
\end{figure}

Here $\kappa$ increases linearly with the Rydberg interaction $V$ up to a certain $V$ ($V = 12$) following the relation $\kappa = 0.5 V + 11$, and beyond that, saturates to a constant value $17$ (Fig. \ref{coefnoteqn}). This is why the curves in the left of Fig\ref{Inequal}(b) almost coincide, unlike the curves in Fig \ref{op}(c) with $N_1=N_2$.

We would like to point out that this scheme of sublattice-specific atom loading can have other utilities as well. For example, if the system is treated using an effective Hamiltonian ( as in Eq. \ref{Hamiltonianmodified}), an alternating $N_1$ and $N_2$ would result in a staggered effective detuning $\Delta_m$ that can have interesting consequences \cite{mukherjee2022periodically, sarkar2023quantum}.

\section{Summary}
In this paper, we investigate the Rydberg excitation dynamics in a dissipative optical lattice, where there are multiple atoms occupying the same site. The focus is on different phases and phase transitions in the system, as the on-site interaction strength is varied. As this on-site interaction can easily be tuned either by applying a Feshbach magnetic field \cite{ohashi2005single, chin2010feshbach}, or by simply changing the lattice parameters \cite{albus2003mixtures}, one can have a precise control over the dynamics and the corresponding phases. 

We analyze the system using two different approaches: (i) solving the Master equation with mean-field approximation in 2D or 3D, assuming a large system size and (ii) numerical solution of an equivalent quantum model in 1D, with a small system size. Although both approaches have certain limitations, they are complementary to each other. Therefore, they can be combined to obtain a holistic picture of the entire system dynamics. 

It is observed that there emerges a density-wave-ordered/ non-uniform state in terms of the Rydberg excitation distribution in a certain window of the on-site interaction. This phase is extremely important in the context of quantum information processing, because structures such as this can pave the way for efficient entanglement in the system \cite{wilk2010entanglement, carr2013preparation}. Therefore, by changing $U$, one can have an additional handle to generate and control the many-particle entangled state. It is also evident that a selective loading of particles (such that the number of atoms differs in the superatoms belonging to two different sublattices) can lead to a wider and more robust antiferromagnetic region in the phase space. There are two subclasses of this non-uniform phase : antiferromagnetic and oscillatory that emerge in both equally populated and unequally populated sublattices. The origin of these phases is probed both using a fixed point calculation, and a study of correlation using a semi-classical Monte Carlo method. 

We also study the scaling forms and criticality associated with the antiferromagnetic transition, where $U$ is the driving parameter. It follows that, although conventional critical exponents are not found, the order parameter can be expressed in a scale-free form using certain scaling exponents. These exponents vary linearly with the Rydberg interaction strength, indicating that the system has the signature of weak universality. 

Future works in this direction might involve different protocols for the Rabi coupling $\Omega$, and check whether an explicit time-dependence in its form  can lead to interesting new phases in terms of the Rydberg excitation distribution.

\section{Acknowledgement}
RD would like to acknowledge Science and Engineering Research Board (SERB), Department of Science and Technology, Govt.of India for providing support under the CRG scheme (CRG/2022/007312).

SI would like to acknowledge DST-INSPIRE, Department of Science and Technology, Govt. of India, for providing support under the AORC scheme of the INSPIRE Program ( DST/INSPIRE Fellowship/2020/IF200534).

\bibliography{bibi.bib}

\end{document}